\documentclass[%
aps,
pre,
groupedaddress,
showkeys,
 amsmath,amssymb,
reprint,%
]{revtex4-2}
\usepackage{graphicx}
\usepackage{dcolumn}
\usepackage{bm}
\usepackage{xcolor}

\begin{document}


\title{Identifying the approach of a major earthquake} 
\thanks{To the memory of the Academician Seiya Uyeda.}

\author{Panayiotis A. Varotsos}
\email{pvaro@otenet.gr}
\affiliation{Section of Condensed Matter Physics and Solid Earth Physics Institute, Physics Department, National and Kapodistrian University of Athens, Panepistimiopolis, Zografos 157 84, Athens, Greece}

\author{Nicholas V. Sarlis}
\affiliation{Section of Condensed Matter Physics and Solid Earth Physics Institute, Physics Department, National and Kapodistrian University of Athens, Panepistimiopolis, Zografos 157 84, Athens, Greece}

\author{Toshiyasu Nagao}
\affiliation{Natural Disaster Research Section (NaDiR), Global Center for Asian and Regional Research, University of Shizuoka, 3-6-1,Takajo, Aoi-Ku, Shizuoka-City, Shizuoka, 420-0839,Japan}



\date{\today}

\begin{abstract}
 By analyzing the seismicity in natural time and studying 
 the evolution of the fluctuations of the entropy change
 of seismicity under time reversal for various scales of 
 different length  $i$  (number of events), we can 
 identify the approach of 
a major earthquake (EQ) occurrence. The current investigation  
 is extended from 1984 until now for the seismicity of Japan.
 \end{abstract}
\maketitle 


\section{Introduction}
For a time series comprising $N$ events, we define as natural time $\chi_k$  for the occurrence of the $k$-th event the quantity  $\chi_k=k/N$ \cite{NAT01,NAT02,NAT02A}. 
Hence, we ignore the time intervals between consecutive events, but preserve 
their order and energy $Q_k$. 
The evolution of the pair $(\chi_k,p_k)$ is studied, where  
$p_k=Q_k/\sum_{n=1}^N Q_n$
is the normalized energy for the  $k$-th event.
The  entropy $S$ in natural time is given by \cite{NAT03B}
\begin{equation}\label{eq3}
S=\langle \chi \ln \chi \rangle - \langle \chi \rangle \ln \langle \chi \rangle,
\end{equation} 
where $\langle f(\chi) \rangle=\sum_{k=1}^N p_k f(\chi_k)$ denotes the average value of $f(\chi)$ weighted by $p_k$, i.e.,
$\langle \chi \ln \chi \rangle = \sum_{k=1}^N p_k (k/N) \ln (k/N)$ and $\langle \chi \rangle = \sum_{k=1}^N p_k (k/N) $.
 Upon considering \cite{NAT05B,SPRINGER} the time-reversal $\hat{T}$, i.e., $\hat{T}p_k=p_{N-k+1}$, the entropy obtained by Eq. (\ref{eq3}), labelled by $S_-$, is given by  
\begin{eqnarray}\label{defS2}
S_-=\sum_{k=1}^N p_{N-k+1} \frac{k}{N} \ln \left( \frac{k}{N} \right)- \nonumber \\
\left(\sum_{k=1}^N p_{N-k+1} \frac{k}{N} \right) \ln \left( \sum_{k=1}^N p_{N-k+1} \frac{k}{N} \right), 
\end{eqnarray}
which  is different from $S$. Thus, a change  $\Delta S \equiv S -S_-$ in natural
 time under time reversal emerges and 
 the fluctuations of the entropy change $\Delta S_i$ under time reversal are obtained through the measure $\Lambda_i$
\begin{equation}\label{eq5}
\Lambda_i = \frac{\sigma (\Delta S_i)}{\sigma (\Delta S_{100})}
\end{equation}
where $\sigma (\Delta S_i)$  is the standard deviation  of the time series of $\Delta S_i \equiv S_i -(S_-)_i$ and 
 the denominator stands for the standard 
 deviation $\sigma (\Delta S_{100})$ of the time series of $\Delta S_i$ of 
 $i$=100 events.  
 For physical reasons explained 
 in Ref.\cite{VAR24SCIREP},  the computation started from 1 January 1984 for the scales 
 $i=$2000, 3000, and 4000 events. We clarify that 
 the minimum of $\Delta S$ being accompanied by an evident 
 increase in the measure $\Lambda_i$ can be uniquely distinguished 
 as EQ precursor \cite{VAR23}.

\begin{figure}[h]
\includegraphics[scale=0.35]{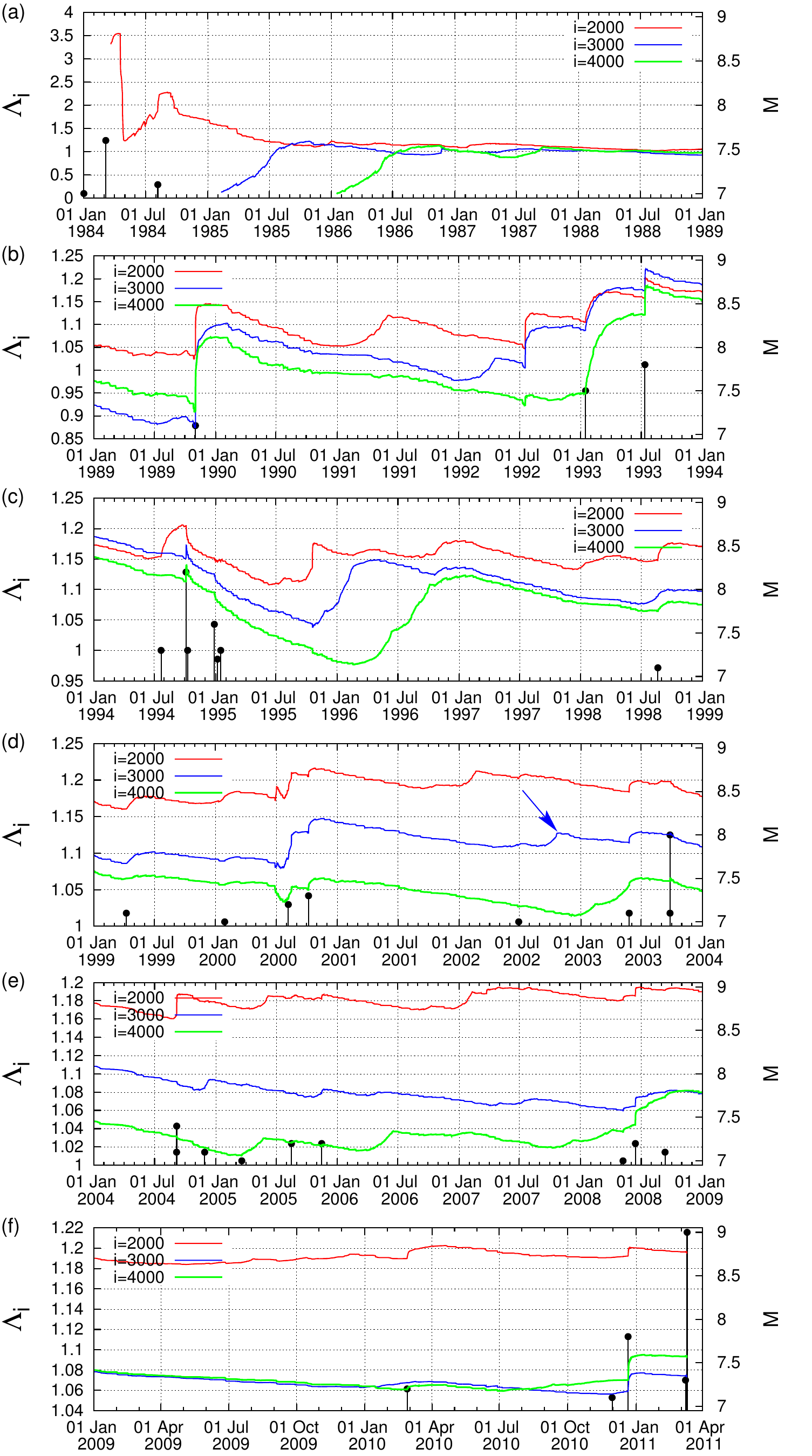}
\caption{{The}  complexity measure $ \Lambda_i $ versus 
the conventional time for the scales $i$ = 2000 (red), 
3000 (blue) and 4000 (green) events  since 
1 January 1984 until the $M$9 Tohoku EQ  on {11 March 2011}
for appropriate vertical scales in each panel. The black solid circles
 show the magnitudes ($M \geq7 $) of EQs{, which are} read in the right scale.
 Taken from {Ref.} 
~\cite{ENTROPY18}. In panel (d), the scale 3000 events 
exhibits a maximum in the range from 1 July 2002 to 1 January 2003 while the other 
two scales 2000 and 4000 events gradually decrease.  Some months later, i.e., on 
26 September 2003 the Tokachi-Oki EQ of magnitude
 $M$8.0 occurred.  \label{f1} }\end{figure}

\begin{figure}[h]
\includegraphics[scale=0.35,angle=0]{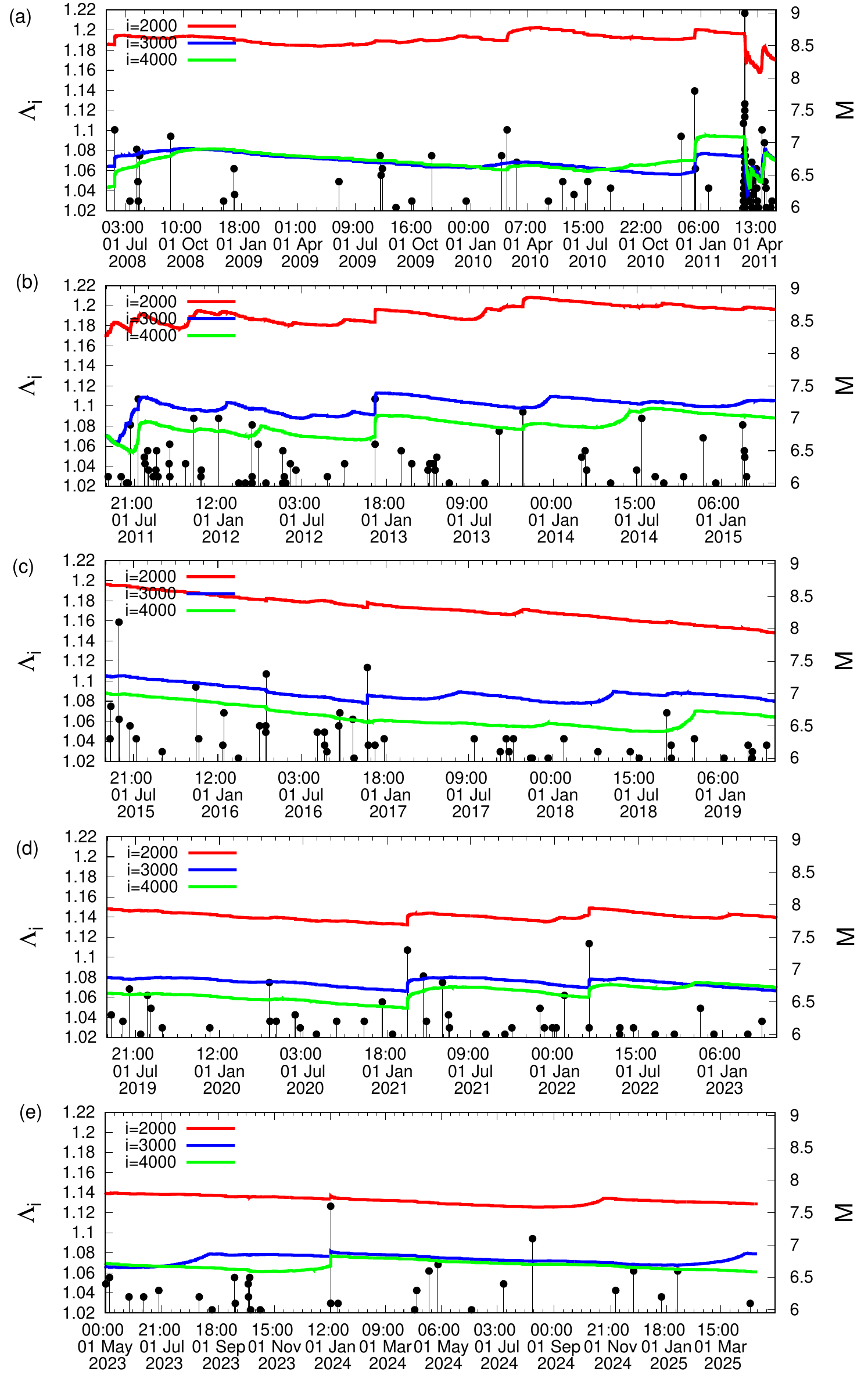}
\caption{The measure $\Lambda_i$ 
versus the conventional time for the scales $i=$2000 (red), 
3000 (blue), and 4000 (green) events from
 1 July 2008 until 10 April 2025.}\label{f2}
\end{figure}

 \section{The $M$9 Tohoku EQ on 11 March 2011}
 
As already mentioned in Ref. \cite{VAR24SCIREP}
during almost a decade from 1 January 1998 until 14 June 2008, 
the date at which 
a $M$7.2 EQ occurred, 
there  exists  no intersection between the 
curves of the aforementioned three scales since the scale $i=2000$ events lies in 
the highest level, the scale $i=3000$ events in the 
middle level and the scale $i=4000$ events in the lowest level
 (see Fig.\ref{f1}). 
Approximately, from the latter date the curve of the 
scale $i=4000$ events shows a clear increase, 
thus finally almost overlapping the curve of the scale $i=3000$ events 
until almost 5 August 2010. From thereon, however, the curve 
corresponding to $i=4000$ events exceeds the one of 3000 events 
(cf. at this date the two curves intersect) and subsequently it
exhibits an abrupt increase upon the 
occurrence of the $M$7.8 
EQ on 22 December 2010 in the Izu-Ogasawara Arc
 at 27.05$^o$N 143.94$^o$E, 
which constitutes an evident intersection.
{Almost 2$\frac{1}{2}$  months after the completion of 
this intersection the $M$9 EQ occurred on 11 March 2011{, cf. 
for other precursory phenomena see \cite{PNAS13,SKO14,VAR17,angeo}.} 

{Remarkably,} on this date (22 December 2010) 
of the abrupt increase of $\Lambda_i$ {the following} additional fact 
{is} observed: The abrupt 
increase conforms to the 
seminal work by Lifshitz and Slyozov \cite{LIF1961} and 
independently by Wagner \cite{WAG1961} (LSW) for phase
 transitions showing that the characteristic size of the 
 minority phase droplets exhibits a scaling behavior 
 in which time ($t$) growth has 
 the form $A (t - t_0)^{1/3}$. 
 It was found that the
 increase $\Delta \Lambda_i $ of $ \Lambda_i $ follows
 the latter form and that the prefactors $A$ are 
 proportional to the scale $i$, while the exponent $(1/3)$ 
is independent of $i$ \cite{ENTROPY18}.

Let us summarize: The scale $i=4000$ events started to overlap the scale $i=3000$ events just after the 
date 14 June 2008 and the major $M$9 Tohoku EQ occurred on 11 March 2011, i.e., somewhat less than around 
3 years. 

\section{The $M$7.6 Noto EQ on 1 January 2024} 
As described in Ref.\cite{VAR23J}, the scale 4000 
events started to overlap the scale $i$=3000 events 
on 27 October 2022
until 27 June 2023. The $M$7.6 Noto EQ occurred on 1 January 2024, i.e., 
appreciably shorter than the period of around 
3 years mentioned above in the case of the $M$9 Tohoku EQ. 
The feature of the evolution of the measure 
  $ \Lambda_i $ depicted in Figs. \ref{f1}(f) and \ref{f2}(a) 
  for the $M$9 Tohoku EQ are similar but 
  differ markedly from the feature of 
  Figs.\ref{f2}(e) and \ref{fnew}{(a)}  of the current case
 in the following sense: During the {vast majority of the period depicted in Fig.\ref{f2}(e)}, the scale 
3000 events exceeds markedly the scale $i=4000$ events, 
while in the case of $M$9 Tohoku EQ the curve 
for the scale {$i=3000$} events exceeded for some months
 clearly the curve of the scale {$i=4000$} events.

\begin{figure}
\includegraphics[scale=0.55,angle=0]{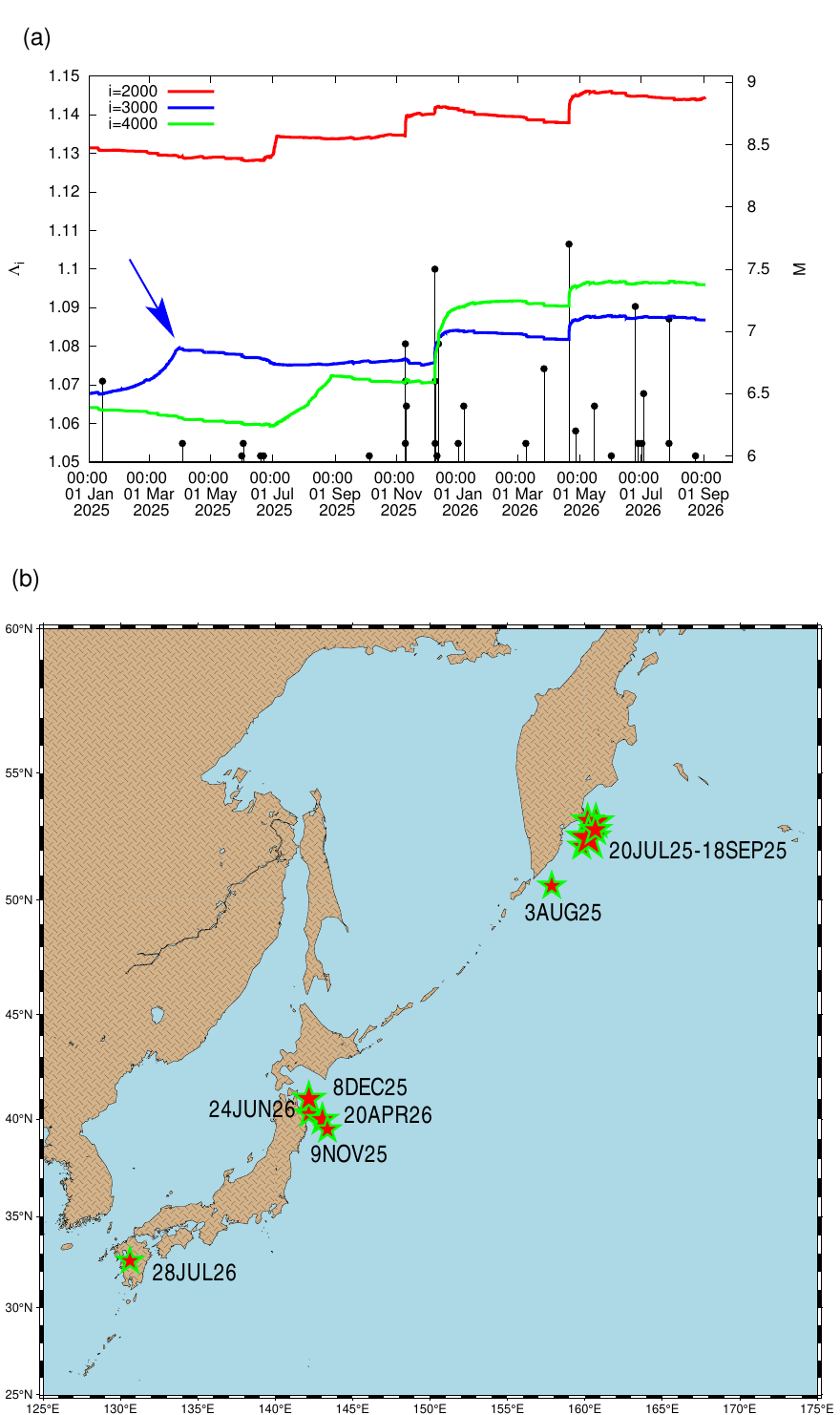}
\caption{(a):The measure $\Lambda_i$ 
versus the conventional time for the scales $i=$2000 (red), 
3000 (blue), and 4000 (green) events from
 1 January 2025 until 3 September 2026. (b):Map of Japan and the surrounding area depicting the epicenters and time of occurrence of the strongest EQs 
 depicted in (a). The recent seismic activity close to Kamchatka Peninsula is also depicted. The data from the NEIC-PDE catalog available from the 
 United States Geological Survey (USGS).  }\label{fnew}
\end{figure}

{\section{{the seismic activity after 1 January 2025 and the complexity measures}}}

{In the third  version of this preprint
 (submitted on 7 May 2025), we drew attention to Fig.\ref{f1}(d),  
 where the 
 $\Lambda_{3000}$  values exhibited between 
 the range 1 July 2002 to 1 January 2003  a considerable 
 increase (indicated by the blue arrow)
 while the curves  of the scales 2000 events and 4000 events
 gradually decrease and some months 
later, i.e., on 26 September 2003 the
 Tokachi-Oki EQ of magnitude $M$8.0  occurred. This behavior is similar to
 the one observed during the range 1 March to 1 May 2025 indicated 
 in  Fig.\ref{fnew}(a) by the blue arrow. This was followed 9 months later
 on 8 December 2025 at 14:15 UTC by  
 an EQ of magnitude $M$7.5 at 41.043$^o$N 142.141$^o$E
 close to Aomori Prefecture, Japan, see Fig.\ref{fnew}(b). 
 Moreover, on 20 April 2026 at  07:53 UTC an EQ of magnitude $M$7.7 occurred 
 at 39.971$^o$N 143.059$^o$E, see Fig.\ref{fnew}(b).}
 
 {Interestingly, during the period between the $\Lambda_{3000}$ maximum 
indicated by the arrow in Fig.\ref{fnew}(a) and the two aforementioned EQs in Japan,
a very strong EQ activity was observed  almost 14$^o$ NE from the upper NE 
corner of our study area at Kamchatka Peninsula, see Fig.\ref{fnew}(b).}

\begin{figure}
\includegraphics[scale=0.35,angle=0]{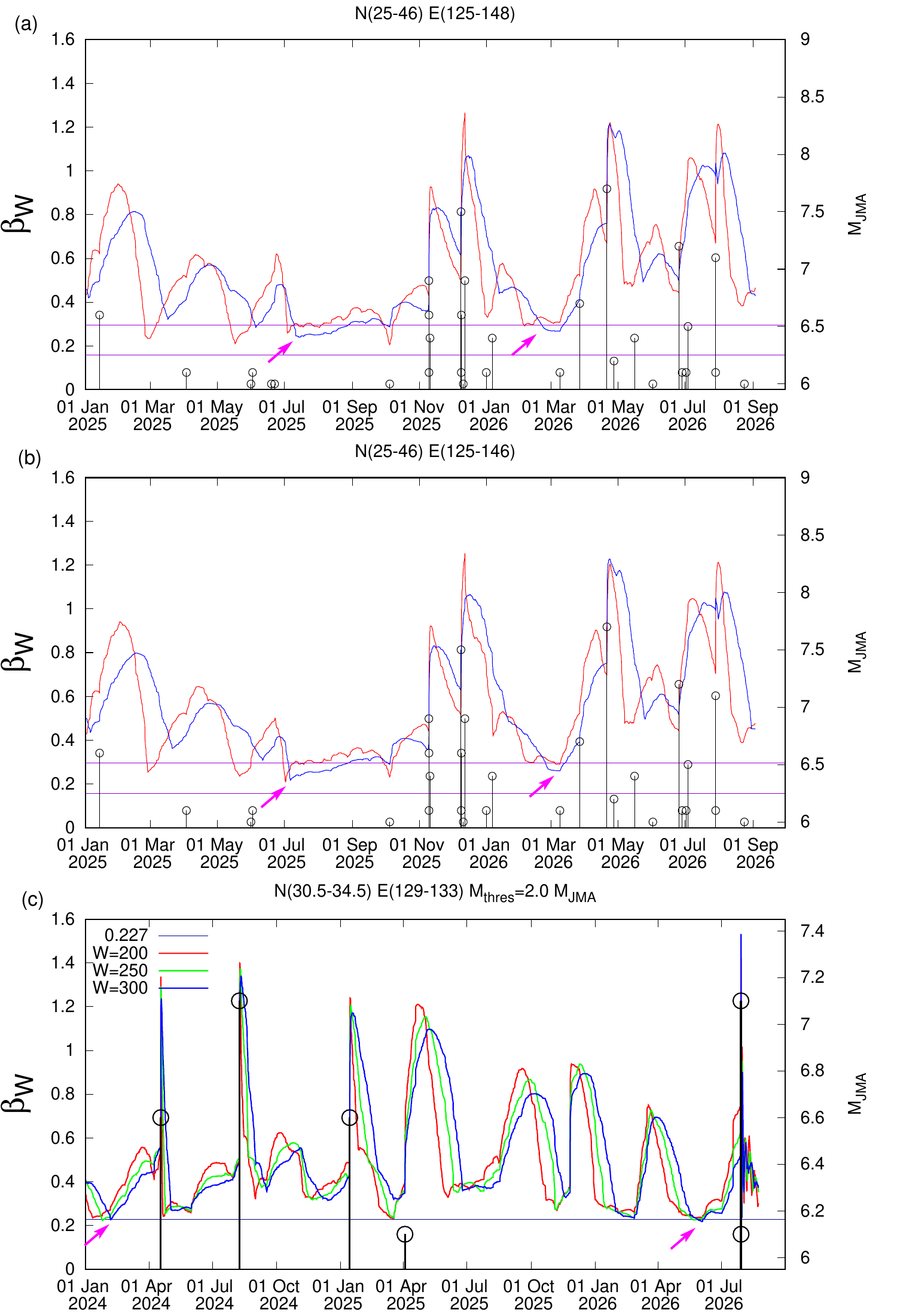}
\caption{Study of the variability $\beta_W$ of the order
 parameter of seismicity $\kappa_1$ 
in natural time for various areas of Japan. 
}
 \label{fKuma}
\end{figure}

{Focusing on the evolving $\Lambda_i$ 
 feature of Fig.\ref{f2}(e), it has not yet reached
 an almost 
 3 year period of the case of the $M$9 Tohoku EQ 
 (see Fig.\ref{f1}f) exhibiting 
 considerable overlapping of the scale 4000 
 events on the scale 3000 events.} In Fig.4 of 
 Ref.\cite{VAR24SCIREP}, only for a period
 of around 7 months (i.e., from 27 October 2022 to until 
 27 June 2023) such an overlapping appeared. Upon the occurrence of 
 the $M$6.9 EQ on 12 December 2025 such an overlapping again appeared
and  $\Lambda_{4000}$ exceeded  $\Lambda_{3000}$ as shown in Fig.\ref{fnew}(a).
In particular, on 8 December 2025, the aforementioned $M$7.5 EQ occurred, 
at the same time, $\Lambda_{4000}$ rapidly increased, 
and the intersection with the $\Lambda_{3000}$ curve was observed 
with the occurrence of the $M$6.9 EQ on December 12th.
Let us wait a few or more months to see the evolution of $\Lambda_i$.

{As pointed out in Ref. \cite{VAR24SCIREP},  
studying the evolution of $\Lambda_i$ curves, 
we obtain a procedure that identifies when the accumulation of stress 
occurs well before major EQs. The aforementioned 
behavior indicated by the arrows in Figs. \ref{f1}(d) and \ref{fnew}(a), 
to be labeled therefrom by the symbols 
'$\backslash \Lambda  \backslash$', precede with a 
lead time of 4 to 17 months all 
major EQs of magnitude $M$ in the range 7.6 to 8.1 which are unrelated 
to stronger EQs. Under this assumption there are only 6  (13-month) periods 
to be identified in the totality of 500 months studied (hence the probability of 
success is $6 \times 13/500=15.7$\%) and 5 of these 6 periods were preceded by
$\backslash \Lambda  \backslash$ (cf. the ultra-deep 2015 Ogasawara EQ, see \cite{VAR21},  
was also preceded by $\backslash \Lambda  \backslash$ 
during December 2013). Additionally, there are 3 more $\backslash \Lambda  \backslash$
 which were followed 
by smaller EQs. A simple binomial distribution calculation reveals that 
the probability to hit at least 5 periods of interest by chance when making
 8 attempts, i.e., the $p$-value,  is less 
than 0.4\%.  }
 
{Since the study of $\Lambda_i$ curves has 
a lead time ranging from 4 to 17 months, for a better estimation of the time 
of occurrence of the impending EQ we can resort to the minima of the fluctuations 
$\beta_W$ of the order parameter $\kappa_1$ of seismicity in natural time which 
have a smaller lead time \cite{PNAS13,PNAS15,SPRINGER23}. 
The details of the definition of $\beta_W$ can be found 
in Refs.\cite{SPRINGER,PNAS13,SPRINGER23} while some of 
their applications for EQ prediction in \cite{JGR14,Chaos15,PER21}.
Figure \ref{fKuma} is a summary figure of the $\beta_W$  minima analysis preceding 
the strong EQs depicted in Fig.\ref{fnew}(a). 
In particular, panels (a) and (b) correspond to the areas analyzed in Ref.\cite{JGR14}, 
while panel (c) to an analysis within the region 
N$_{30.5^o}^{34.5^o}$E$_{129^o}^{133^o}$ surrounding Kyushu, 
where the deadly Kumamoto EQ took place on 28 July 2026.
The horizontal lines have been drawn 
as a guide to the eye. The magenta arrows indicate the variability minima
before the 
stronger EQs that can be identified by imposing appropriate conditions on $\beta_W$ 
in a fashion similar to Refs.\cite{PNAS13,JGR14}. We observe that the indicated 
variability minima precede the strong EQs in each case with    lead times significantly
 smaller  than those of the corresponding $\backslash \Lambda  \backslash$ precursors.
}

\section{Main conclusion}
 
 {By analyzing the seismicity of Japan in natural time and studying 
 the evolution of the fluctuations of the entropy change
 of seismicity under time reversal for various scales of 
 different length  $i$  (number of events), we found a new characteristic behavior, 
labeled $\backslash \Lambda  \backslash$, 
 of the complexity measures $\Lambda_i$. The $\backslash \Lambda  \backslash$
behavior indicated by the arrows in
 Figs. \ref{f1}(d) and \ref{fnew}(a) precedes EQs of magnitude $M$ 
 in the range 7.6 to 8.1 with a lead time of 4 to 17 months. Such a behaviour 
 was identified  before the occurrence of the strongest EQs in Japan since 
 11 March 2011 the day 
 the $M$9 Tohoku EQ 
  (cf. the EQ on 1 January 2024 was identified by a  previous  
 preprint of ours, see \cite{VAR23J}, and was also preceded by a 
 $\backslash \Lambda  \backslash$ during August 2023).}

\providecommand{\noopsort}[1]{}\providecommand{\singleletter}[1]{#1}%

\end{document}